\documentclass{moriond}
\usepackage{xspace}
\usepackage{subfig}
\usepackage[permil]{overpic}
\usepackage{amsmath}
\usepackage{lineno}

\def\be{\begin{equation}}
\def\ee{\end{equation}}
\def\bea{\begin{eqnarray}}
\def\eea{\end{eqnarray}}

\newcommand{\Photo}{\includegraphics[height=35mm]{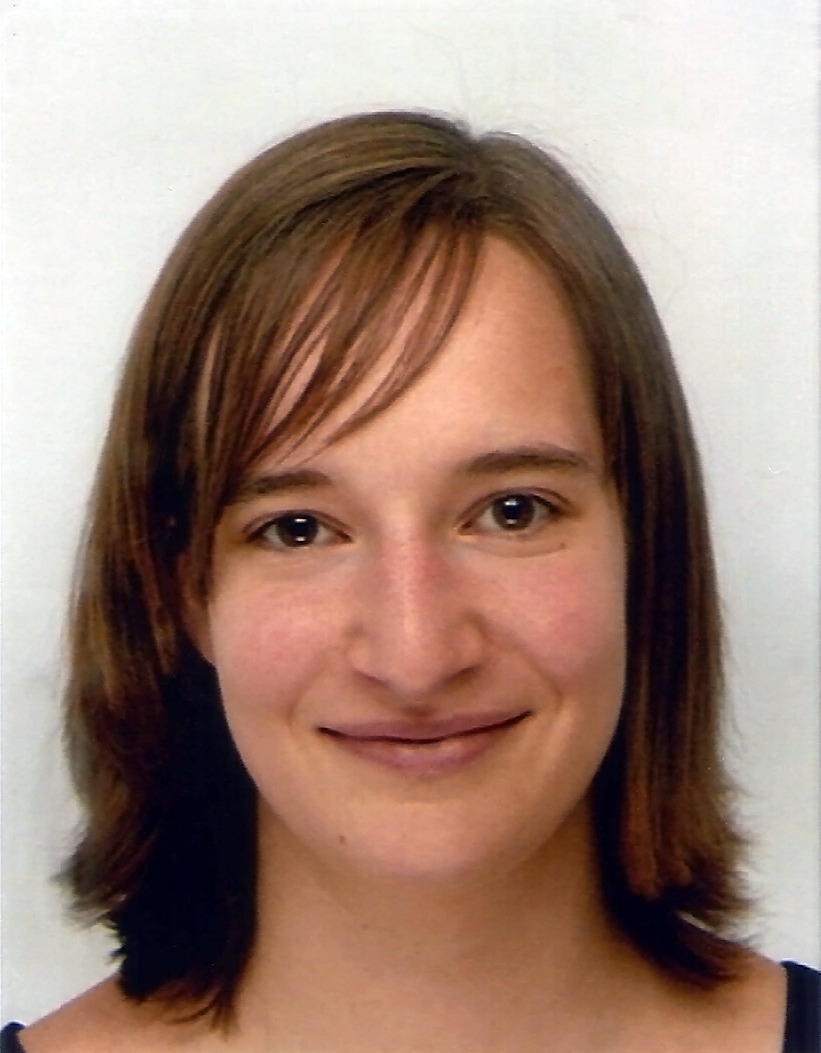}}

\newcommand{\pT}           {$p_{\,\rm T}$\xspace}
\newcommand{\Dphi}         {$\Delta\varphi$\xspace} 
\newcommand{\Deta}         {$\Delta\eta$\xspace}

\newcommand{\GeVc}         {~GeV/\textit{c}\xspace}
\newcommand{\snn}          {\mbox{$\sqrt{s_{\rm NN}}=2.76$~TeV}\xspace}

\pdfoutput=1

\begin{document}
\vspace*{4cm}
\title{Anomalous evolution of the near-side jet peak shape\\ in Pb-Pb collisions with ALICE}

\author{MONIKA KOFARAGO  for the ALICE collaboration}

\address{MTA Wigner RCP, Hungary 1525 Budapest Pf. 49.}

\maketitle\abstracts{
Two-particle angular correlations are sensitive probes to study the interaction of jets with the flowing medium produced in heavy-ion collisions. These interactions may appear as modifications of the near-side jet peak compared to pp collisions. In these measurements, the associated per-trigger yield is calculated from the relative azimuthal angle and pseudorapidity between a trigger particle with higher $p_{\rm T}$ (1\GeVc$ < p_{\rm T} < 8$ \GeVc) and an associated particle. Subsequently, the near-side peak width and shape are extracted as a function of $p_{\rm T}$ and centrality. Results obtained by the ALICE detector from Pb--Pb and pp collisions are presented. In Pb--Pb collisions, a significant broadening of the peak in central events at low $p_{\rm T}$ is observed in the data, and is more pronounced in the $\Delta\eta$ direction than in the $\Delta\varphi$ direction. A novel feature is also observed at low $p_{\rm T}$ in central events: the peak departs from the Gaussian shape, and a depletion around its center appears. To put the broadening and the depletion in context with the strength of longitudinal, radial and elliptic flow, the results are compared to AMPT simulations, which suggest that radial and longitudinal flow play a significant role in the appearance of the observed features.}

\section{Motivation}
In the collisions of heavy-ions, a hot and dense medium, the so called quark-gluon plasma, is created. Jet pairs produced from hard scatterings occurring during the collision, traverse this expanding medium. Jets lose energy in the medium by induced gluon radiation and elastic scatterings, which can result in jet pairs with highly asymmetric energies~\cite{Aad:2010bu,Chatrchyan:2011sx}.

The reconstruction of jets below a certain transverse momentum (\pT) becomes problematic due to the large fluctuating background in heavy-ion collisions. Instead, at low \pT, angular correlation measurements can be used to study the interaction of jets with the produced medium. In these measurements, the azimuthal angle (\Dphi) and the pseudorapidity (\Deta) difference between particle pairs are calculated. Back to back jets manifest themselves as a peak around $(\Delta\varphi,\Delta\eta) = (0,0)$ and as an elongated object in \Deta at \Dphi = $\pi$. The interaction of jets with the medium causes a change in the shape of this peak compared to the shape in pp collisions~\cite{Armesto:2004pt,Armesto:2004vz,Romatschke:2006bb}. Such a modification has been seen by the STAR collaboration in central Au--Au collisions at $\sqrt{s_ {\rm NN}} = 200$~GeV~\cite{Agakishiev:2011st}. This paper extends these measurements in energy and \pT reach by the ALICE detector at the LHC. The full details of the presented analysis are described elsewhere~\cite{Adam:2016ckp,Adam:2016tsv}.

\section{Analysis}
For the current analysis, Pb--Pb and pp events taken at \snn by the ALICE detector are used~\cite{Aamodt:2008zz}. To build two-particle correlations, a trigger particle from a certain \pT window within 1\GeVc$<p_{\rm T,trig}<8$\GeVc and an associated particle from a certain \pT window also within 1\GeVc$<p_{\rm T,assoc}<8$\GeVc are taken. The \pT window of the associated particle can be either lower or the same as for the trigger particle in which case only pairs with $p_{\rm T,assoc}<p_{\rm T,trig}$ are used to avoid double counting. From these pairs the following per trigger yield histograms are built:
\begin{equation}
  \frac{1}{N_{trig}}\frac{\text{d}^2N_{assoc}}{\text{d}\Delta\varphi \text{d}\Delta\eta} = \frac{S(\Delta\varphi,\Delta\eta)}{M(\Delta\varphi,\Delta\eta)}
\end{equation}
where $S(\Delta\varphi,\Delta\eta)$ is the sibling histogram, originating from particles taken from the same event, while the division by $M(\Delta\varphi,\Delta\eta)$ accounts for the limited detector acceptance and for pair inefficiencies in the detector. $M(\Delta\varphi,\Delta\eta)$ is constructed from particle pairs, where the two particles are originating from different events, and it is normalized such that it is unity at $(\Delta\varphi,\Delta\eta) = (0,0)$. Several selection criteria and corrections were applied to the per trigger yield histograms, which are described elsewhere~\cite{Adam:2016ckp,Adam:2016tsv}.

\begin{figure}[!htbp]
  \makebox[\textwidth][c]{
    \subfloat[]{%
      \begin{overpic}[width=0.32\textwidth]{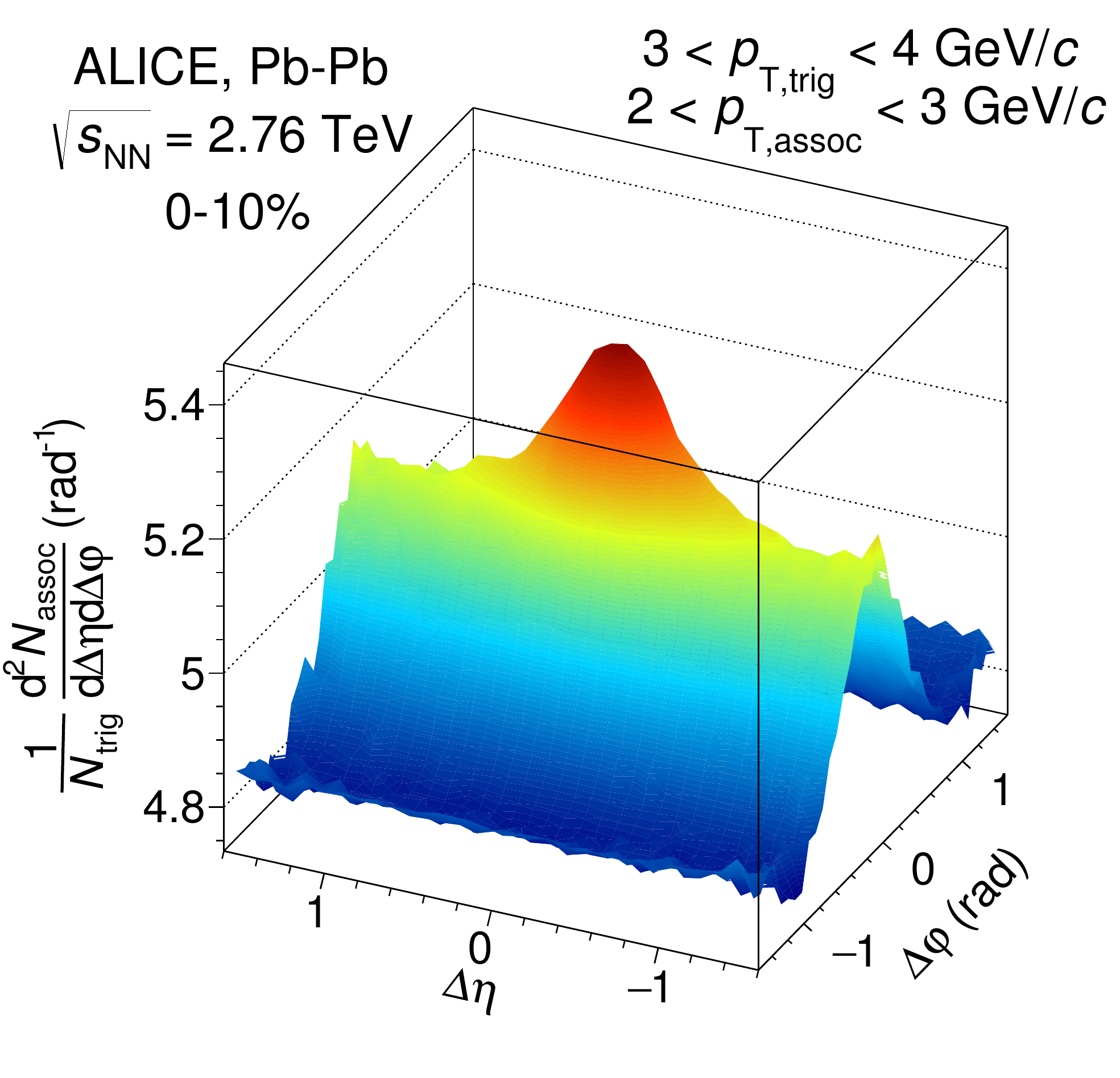}
        \label{subref:fit1}
      \end{overpic}}
    \subfloat[]{%
      \begin{overpic}[width=0.32\textwidth]{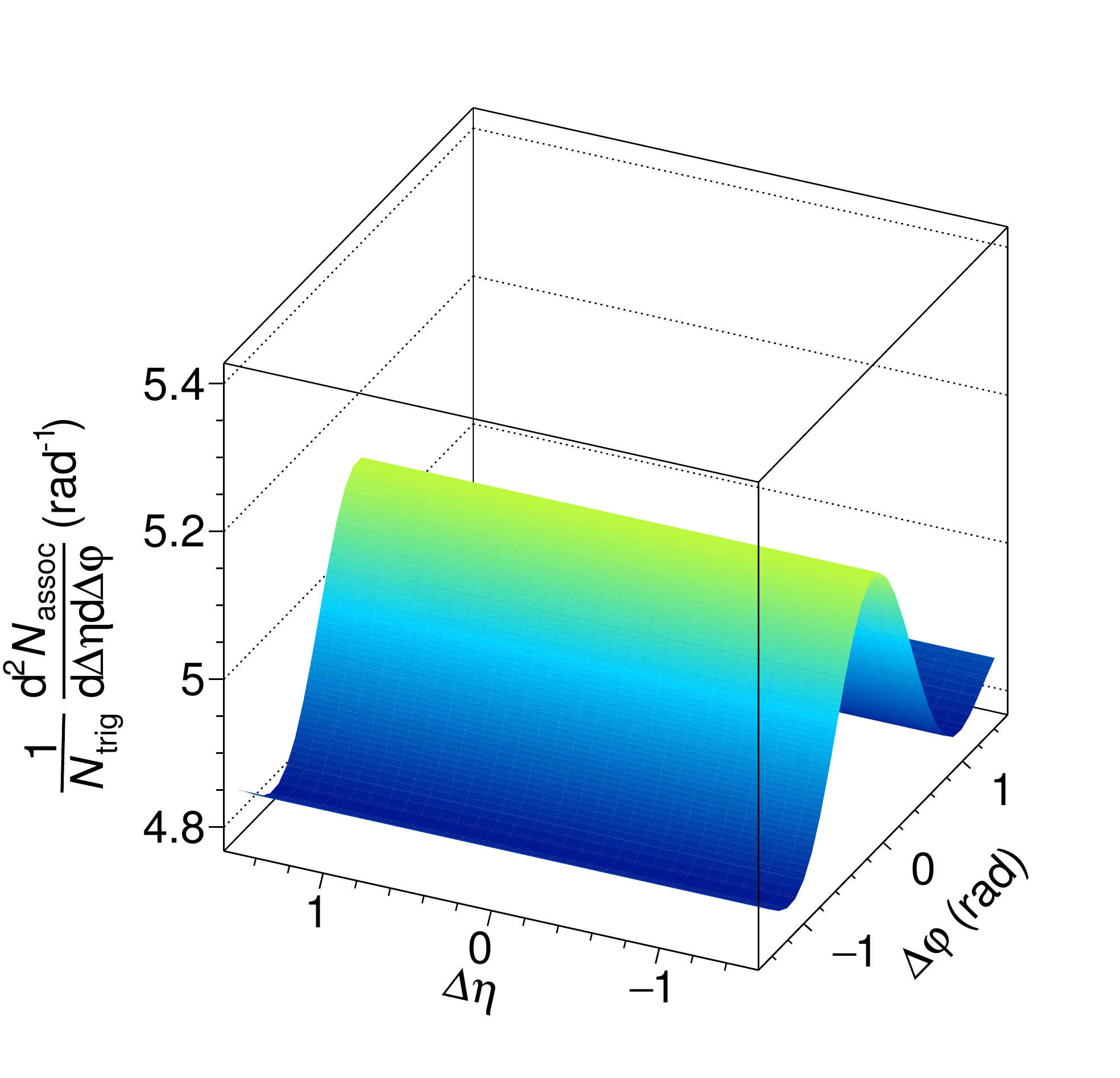}
          \label{subref:fit2}
      \end{overpic}}
    \subfloat[]{%
      \begin{overpic}[width=0.32\textwidth]{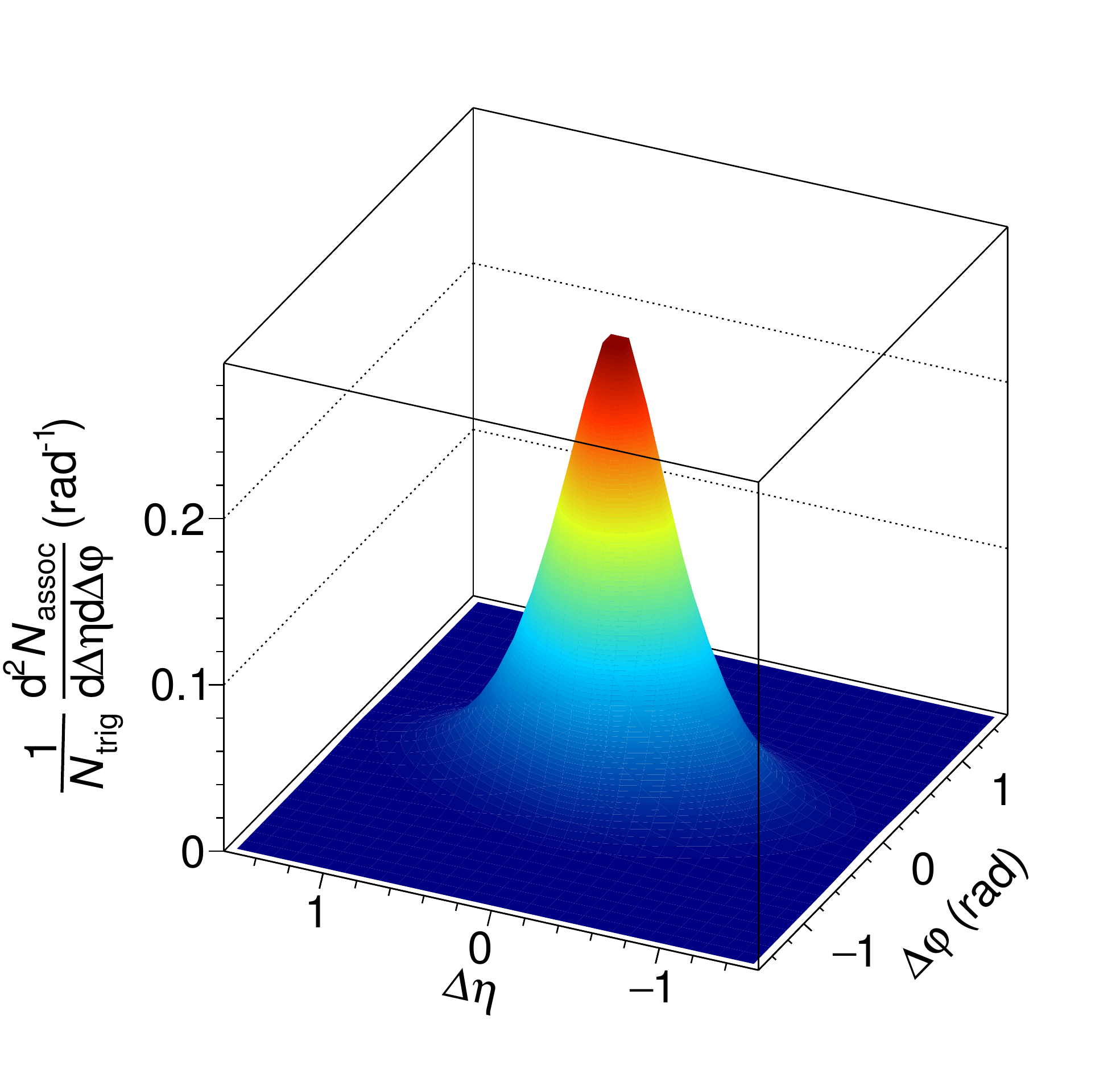}
          \label{subref:fit3}
      \end{overpic}}}
  \caption{An example of the fit of the per trigger yield histograms. Panel \protect\subref{subref:fit1} shows a per trigger yield histogram, panel \protect\subref{subref:fit2} the background part of the fit, while panel \protect\subref{subref:fit3} the signal part.}
  \label{fig:fit}
\end{figure}

These per trigger yield histograms contain the jet signal, but they also contain the background originating from combinatorics and from the flowing medium. To extract the signal, the 2-dimensional histograms are fitted with a function, which describes both the background and the signal:
\begin{equation}
  F(\Delta\varphi,\Delta\eta) = C_1 + \sum_{n=2}^4 2 V_{n\Delta} \cos (n \Delta\varphi) 
    + C_2 \cdot G_{\gamma_{\Delta\varphi},w_{\Delta\varphi}}(\Delta\varphi) \cdot G_{\gamma_{\Delta\eta},w_{\Delta\eta}}(\Delta\eta)
 \label{eq:fit}
\end{equation}
\begin{equation}
  G_{\gamma_x,w_x}(x) = \frac{\gamma_x}{2{w}_x\Gamma (1/\gamma_x)} \exp \left[ -\left(\frac{|x|}{w_x}\right)^{\gamma_x} \right]
\end{equation}
In this function, the background is described by four parameters: $C_1$ is responsible for the combinatorial background, while the $V_n$ parameters describe the background coming from anisotropic flow~\cite{Aamodt:2011by}. The peak is described by a generalized Gaussian ($G_{\gamma_x,w_x}(x)$) function in both the \Dphi and the \Deta directions. The generalized Gaussian has two parameters, one responsible for its width ($w$), and one for its shape ($\gamma$). The shape of the generalized Gaussian function changes between an exponential and a normal Gaussian function in the range $1<\gamma<2$. The fitting of a per trigger yield histogram is illustrated in Fig.~\ref{fig:fit}.

\section{Results}
\begin{figure}[!htbp]
  \makebox[\textwidth][c]{
    \subfloat[]{%
      \begin{overpic}[width=0.45\textwidth]{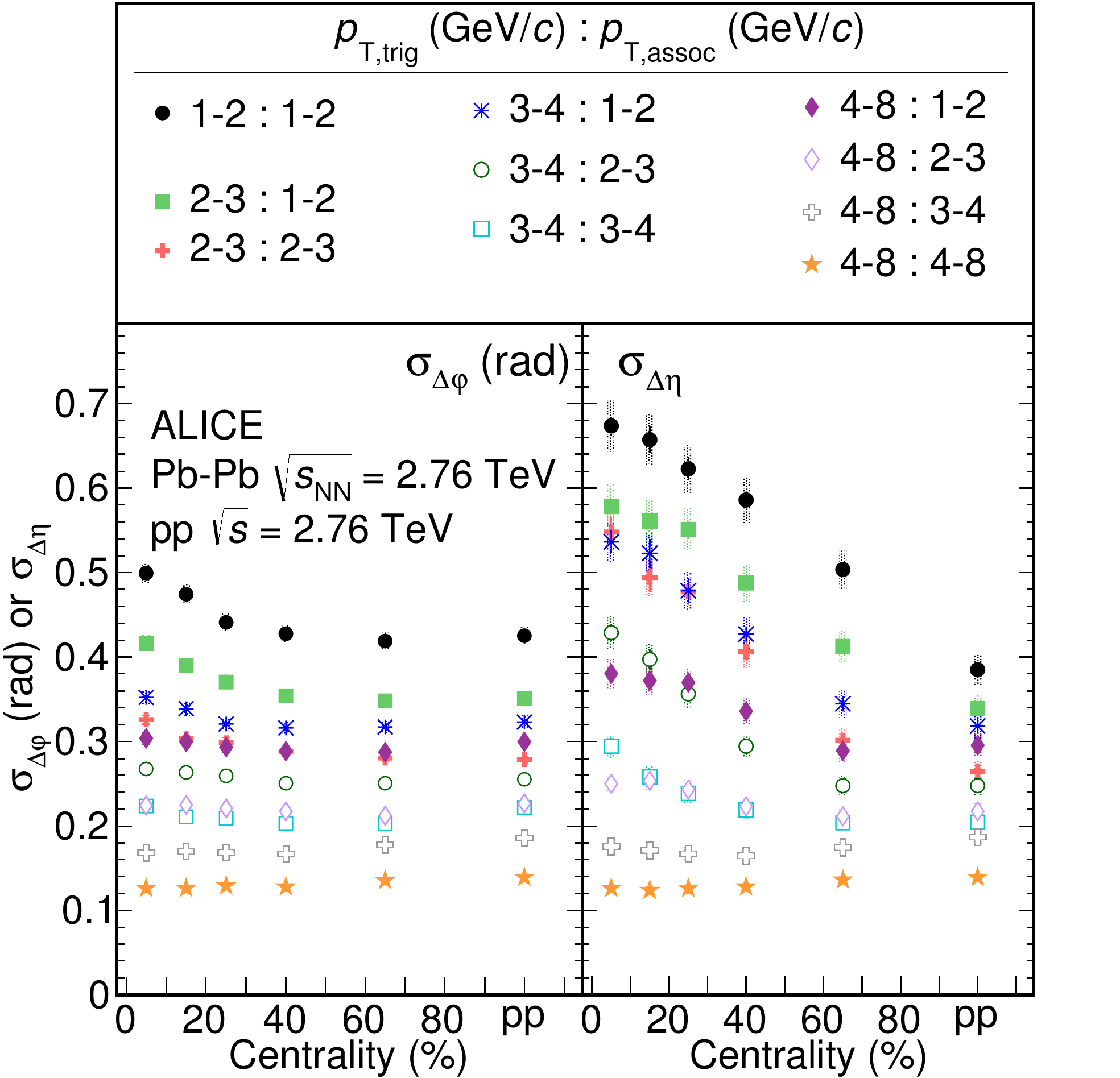}
        \label{subref:widthTwoPanel}
      \end{overpic}}
    \subfloat[]{%
      \begin{overpic}[width=0.45\textwidth]{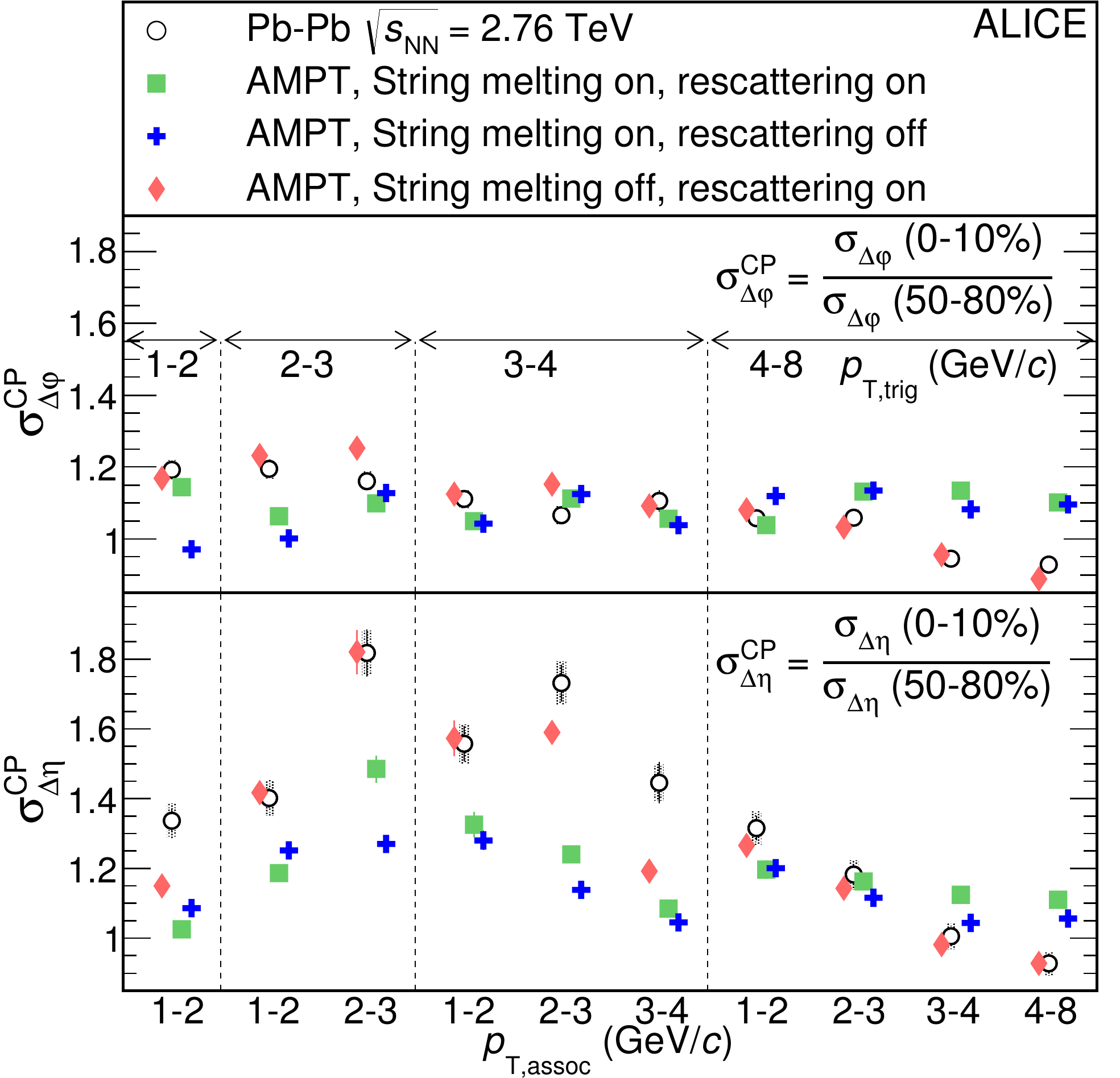}
          \label{subref:CP_with2Panels}
      \end{overpic}}}
  \caption{Panel \protect\subref{subref:widthTwoPanel} shows the width of the jet-peak in Pb--Pb collisions as a function of centrality together with the width in pp (rightmost points) for all measured \pT bins at \snn. Panel \protect\subref{subref:CP_with2Panels} shows the ratio of the width in the most central (0-10\%) and the most peripheral (50--80\%) bins from data, together with simulation results from AMPT.}
  \label{fig:width}
\end{figure}

To describe the peak, its width is calculated as the variance of the generalized Gaussian function. In Fig.~\ref{subref:widthTwoPanel}, this width is shown for both the \Dphi and the \Deta direction from Pb--Pb and pp collisions. At high \pT, where both $p_{\rm T,trig}$ and $p_{\rm T,assoc}$ are above 4\GeVc, the peak is symmetric in both pp and Pb--Pb collisions at all collision centralities. This symmetry disappears at lower \pT in Pb--Pb collisions, while it remains in the case of the pp collisions down to the lowest measured \pT bin (1\GeVc$<p_{\rm T,trig},~p_{\rm T,assoc}<2$\GeVc). In the case of Pb--Pb collisions at lower \pT, the peak is broader in the \Deta direction than in the \Dphi direction.

A broadening towards central events at low \pT is also observed. A small broadening in the \Dphi direction is present, while a more significant broadening can be seen in the \Deta direction, which persists to all except the highest two \pT bins. The increase is quantified in Fig.~\ref{subref:CP_with2Panels} by the ratio of the width in the most central bin and the most peripheral one. It is compared to the same ratio measured in three different settings of generator level AMPT simulations~\cite{Lin:2004en,Xu:2011fi}. The setting with string melting turned off, but hadronic rescattering turned on correctly describes the \pT and centrality evolution of the peak, while the other two settings underestimate the broadening at intermediate \pT in the \Deta direction.

\begin{figure}[!htbp]
  \makebox[\textwidth][c]{
    \subfloat[]{%
      \begin{overpic}[width=0.26\textwidth]{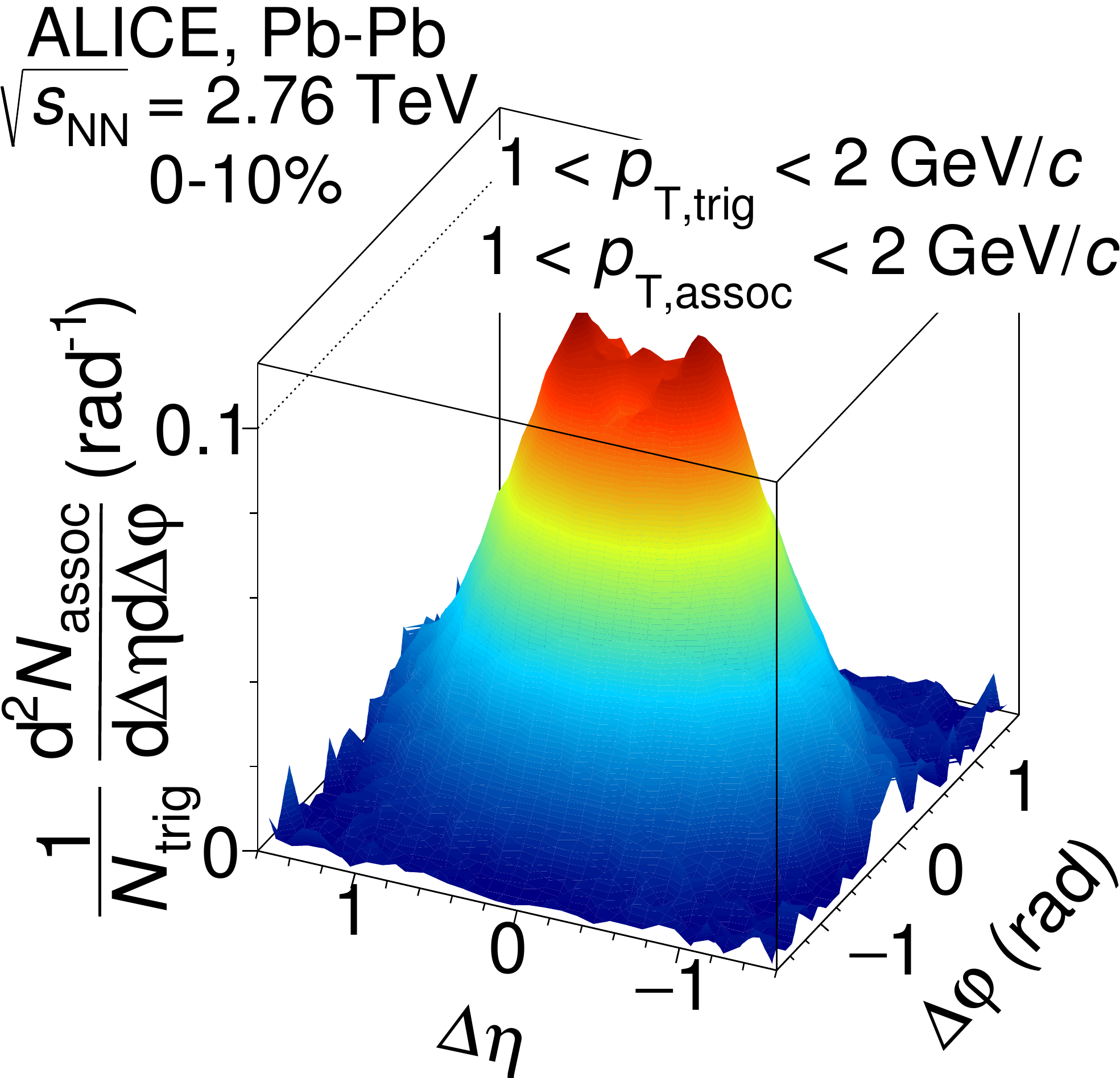}
        \label{subref:results1c_PRL}
      \end{overpic}}\hfill
    \subfloat[]{%
      \begin{overpic}[width=0.53\textwidth]{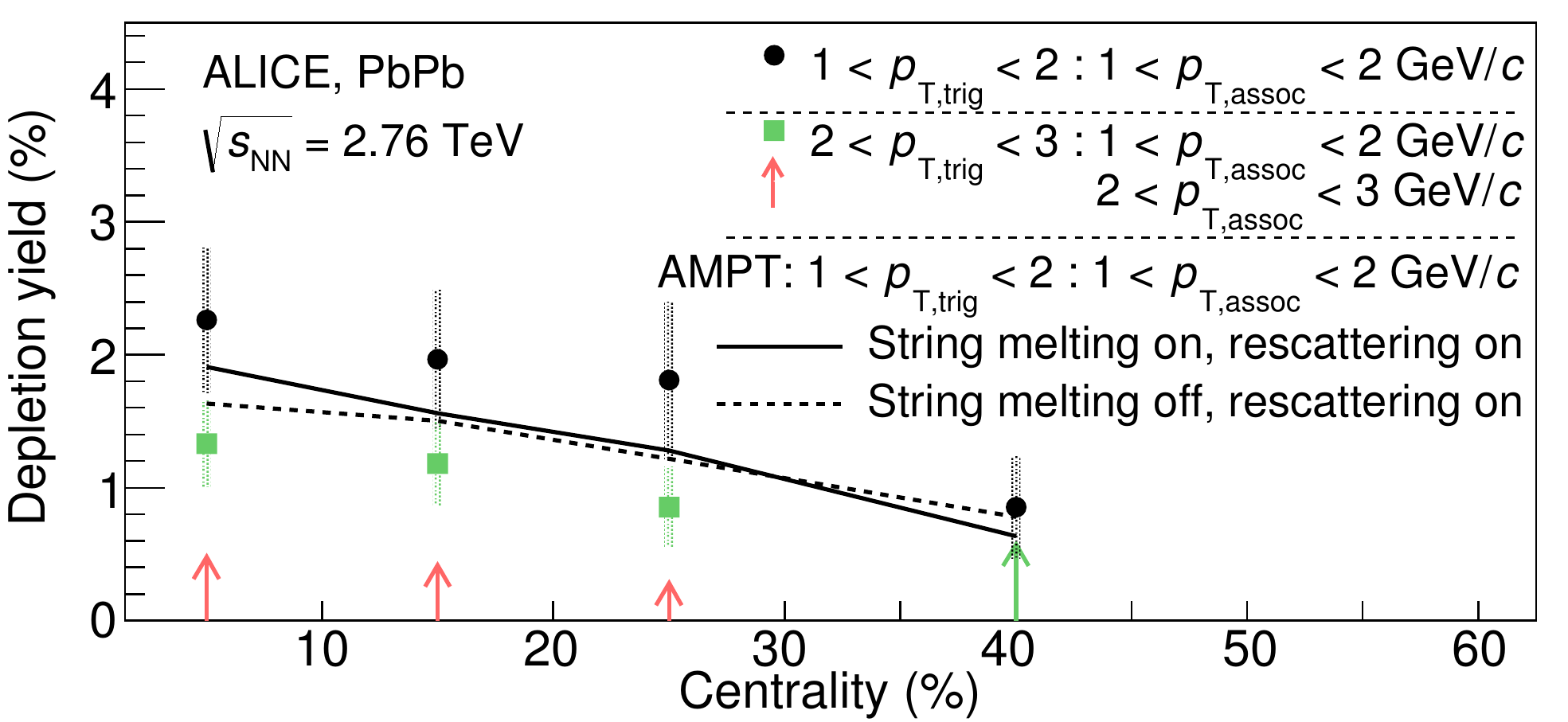}
          \label{subref:depletion_AMPT_comparison_PRL}
      \end{overpic}}}
  \caption{Panel \protect\subref{subref:results1c_PRL} shows an example per trigger yield histogram after the subtraction of the background, while panel \protect\subref{subref:depletion_AMPT_comparison_PRL} shows the depletion yield measured in Pb--Pb collisions at \snn by the ALICE detector (points), compared to simulations form AMPT (lines).}
  \label{fig:dip}
\end{figure}

Apart from the broadening at low \pT a novel depletion around $(\Delta\varphi,\Delta\eta) = (0,0)$ is also seen in the data (Fig.~\ref{subref:results1c_PRL}). The area of this depletion is excluded from the fit to give an unbiased characterization of the width of the peak. The depletion is characterized by the difference of the fit and the histogram in the excluded area, and this difference is normalized to the full yield of the peak. In Fig.~\ref{subref:depletion_AMPT_comparison_PRL}, this depletion yield is overlaid with results from AMPT simulations, which describe the results within the uncertainties of the measurement if hadronic rescattering is turned on, independent of string melting.

\section{Interpretation and conclusions}
To study whether the observations could arise form an interplay of the jets with the flowing medium, the radial flow velocity and the elliptic flow parameter is extracted from both data and the AMPT simulations. The exact values can be found elsewhere~\cite{Adam:2016ckp,Adam:2016tsv}. The best description of the radial flow is given by the AMPT setting with string melting turned off, but hadronic rescattering turned on. This setting describes both the relative evolution of the width and the depletion successfully. The elliptic flow parameter is described by two AMPT settings, when either string melting or hadronic rescattering is turned on. Out of these two, only the latter describes the relative evolution and the depletion properly. From this comparison and from previous work~\cite{Ma:2008nd}, it can be concluded that longitudinal and radial flow are more likely the cause of the presented effects than elliptic flow.

In conclusion, two-particle angular correlation measurements from Pb--Pb and pp collisions at \snn measured by the ALICE detector were shown. A significant broadening in the \Deta direction and a smaller broadening in the \Dphi direction at low \pT towards central events, together with a novel feature, a depletion around $(\Delta\varphi,\Delta\eta) = (0,0)$ were presented. The results were compared to AMPT simulations, which show that radial and longitudinal flow are more likely the cause of the observed effects, as opposed to anisotropic flow. From these measurements we conclude that the broadening and the depletion can be interpreted as an interplay of the traversing jets and the flowing medium.

\section*{Acknowledgments}

This work has been supported by the Hungarian NKFIH/OTKA K120660 grant.

\section*{References}
\bibliography{Kofarago}

\end{document}